# Prediction with Differential Covariate Classification:
## Illustrated by Covariate Classification in Medical Risk Assessment

Atheendar S. Venkataramani (University of Pennsylvania)

Charles F. Manski (Northwestern University)

John Mullahy (University of Wisconsin-Madison)

December 4, 2025

**Abstract:** A common practice in evidence-based decision-making uses estimates of conditional probabilities P(y|x) obtained from research studies to predict outcomes y on the basis of observed covariates x. Given this information, decisions are then based on the predicted outcomes. Researchers commonly assume that the predictors used in the generation of the evidence are the same as those used in applying the evidence: i.e., the meaning of x in the two circumstances is the same. This may not be the case in real-world settings. Across a wide range of settings, ranging from clinical practice to education policy, demographic attributes (e.g., age, race, ethnicity) are often classified differently in research studies than in decision settings. This paper studies identification in such settings. We propose a formal framework for prediction with what we term *differential covariate classification* (DCC). Using this framework, we analyze partial identification of probabilistic predictions and assess how various assumptions influence the identification regions. We apply the findings to a range of settings, focusing mainly on differential classification of individuals' race and ethnicity in clinical medicine. We find that bounds on P(y|x) can be wide, and the information needed to narrow them available only in special cases. These findings highlight an important problem in using evidence in decision making, a problem that has not yet been fully appreciated in debates on classification in public policy and medicine.

Notes: We thank the Editor, Associate Editor, two anonymous referees, seminar participants at the University of Bologna, and Marcella Alsan, Nico Bassols, Federico Crippa, Hanns Kuttner, and Dan Millimet for helpful comments. We gratefully acknowledge Ritikaa Khanna for research assistance. Manski: cfmanski@northwestern.edu; Mullahy: jmullahy@facstaff.wisc.edu; Venkataramani: atheenv@pennmedicine.upenn.edu. The ordering of the authors' names was determined randomly.

What signifies knowing the Names, if you know not the Natures of Things.

Benjamin Franklin, *Poor Richard Improved*, 1750

## 1. Introduction

We consider a decision maker who recommends treatments for members of a heterogeneous population. The decision maker observes certain covariates for each person and wants to make decisions informed by probabilistic predictions of personal outcomes conditional on the observed covariates. The problem is that the available research does not yield the desired conditional predictions. Instead, it yields predictions that condition on covariates whose nature and/or meaning may differ from the covariate information available to the decision maker. The decision maker wants to use the research findings to the extent possible.

This abstract problem has many practical manifestations. A common context is one of a clinician who wants to use the findings of research on medical risk assessment and treatment response to personalize patient care. However, patient covariates are often classified differently in research studies than in clinical practice. Hence, clinicians reading research findings may be uncertain how to interpret them for the patients in their care. Analogous problems exist in education (teachers seeking to personalize instruction for groups of students), management (managers seeking to deploy skill-building interventions among employees), and social policy (a policymaker wishing to tailor interventions to segments of the population to address high rates of poor outcomes such as unemployment).[1]

---

[1] These and other related programs are subjects of a small recent body of economic research, performed contemporaneously with our work. Millimet (2024) studies the implications of different units of geographic classification (e.g., counties and congressional districts) for the estimation of relevant spatial statistics. Closer to the setting we examine, Finlay et al (2024) and Baron et al (2024) investigate implications of differential classification of race and ethnicity in estimating disparities in incarceration rates and foster care placement, respectively.

   This paper's focus on differential covariate classification has commonalities with what is known in some fields (medical informatics, computer science, etc.) as semantic interoperability. One characterization of semantic interoperability is:

In medical contexts — the setting we use to fix ideas — the problem of differential classification can manifest in several different ways. One issue may be different classification of patient health status. Research studies and clinical practice may differ in the procedures and technology used to measure health in physical examinations and in screening or diagnostic tests. Even if there is uniformity in the way data are collected, researchers and clinicians may use different criteria to summarize or categorize patient status. For example, they may use different thresholds to assign patients to risk categories: studies may identify patients with high blood pressure using different cutoffs that may have been standard at different points in time.

Another issue may be different classification of patient demographic attributes. For example, research and practice may differ in their classifications of patient race/ethnicity, where there presently exists no consensus approach. Classification may be based on patient responses to questions asking for self-identification, collection of data on patient ancestry, or by assignments made by an external observer (e.g., by skin tone). Even if there is uniformity in data collection, researchers and clinicians may use different criteria to aggregate the data into racial/ethnic classes, for example using different classification schema to assign race/ethnicity or using broad descriptors (e.g., "Asian") versus sub-categories (e.g., "Asian Indian," "Filipino"). Classification may vary between research and clinical settings for a variety of reasons, including differing incentives, constraints, and norms. The literature sometimes refers to this phenomenon as the fluidity of classification. See Antman and Duncan (2015), Charles and Guryan (2011), Davenport,

---

"Semantic interoperability is the ability of computer systems to exchange data, with unambiguous meaning. It is a requirement not only for health data be shared between different systems or applications, but for them to be understood. *Semantic interoperability refers to the transmission of the meaning of data.*" (PAHO, 2021) (italics added)

That the meaning of or the names attached to a particular phenomenon may vary across circumstances is fundamental to this paper's inquiries.

2020, Penner and Saperstein (2008), Salhi et al., (2024), and Saperstein and Penner (2012), for various perspectives from social science and clinical research.[2]

Whatever the source of *differential covariate classification* (DCC), a consequence is that a clinician who wants to predict patient outcomes conditional on clinically observed patient covariates faces an identification problem. Let y denote a patient's health outcome that has been studied in research and that is of clinical concern. Suppose that a research study measured patient covariates w, enabling researchers to learn and report the conditional probability distribution P(y|w). Suppose that clinical practice classifies patients by covariates x rather than w. Hence, the clinician wants to learn P(y|x) rather than P(y|w). Then the clinical identification problem is to draw conclusions about P(y|x) by combining knowledge of P(y|w) with other information that may be available.

We consider the case of a clinician who treats patients drawn from a population assumed to have the same composition as the one studied in the research. We maintain this assumption to focus attention on the DCC problem. In reality, a clinician may face a further problem of extrapolation (external validity) that arises if the population studied in research differs from the one treated in clinical practice.[3]

The identification problem arising because of DCC is too broad to study effectively in generality, but progress is possible if one considers specific settings, which are common and consequential in the real world. We focus on differential racial/ethnic classification, for two reasons. Racial/ethnic classification is of great current concern in the medical community. Two prominent controversies are whether race and

---

[2] Beyond racial/ethnic classification, other common examples of differential demographic classification include gender, disability status, and family history of illness.

[3] One should not conflate the DCC problem with out-of-population prediction. In the latter case, one observes a distribution $P_0(y|x)$ in population 0 but wants to know the distribution $P_1(y|x)$ in a different population 1. The variables (y, x) are measured the same way in both populations. What differs is their distribution in each population. Out-of-population prediction is an issue of external validity. External validity concerns are sometimes subclassified as concerns about *generalizability* and concerns about *transportability* (see Degtiar and Rose, 2023). We are not aware of these literatures addressing this paper's main concerns. See Manski (2019, chap. 2), on external validity.

The DCC and external validity problems jointly exist if research is conducted in a population described by $P_0(y, x, w)$, where only $P_0(y|w)$ is known, and the clinician delivers care in a population described by $P_1(y, x, w)$. We assume $P_0(y, x, w) = P_1(y, x, w)$.

ethnicity should be conceptualized in ancestral terms, as a social construct, or in some hybrid manner, and whether clinical decision making should consider patients' race and ethnicity however they may be classified (Manski et al., 2023). Classification of race and ethnicity in federal data systems is also a topic of current policy interest, with the White House Office of Management and Budget (OMB) in 2024 publishing a set of revisions to prior classification guidelines known as Statistical Policy Directive 15, or SPD-15.[4] Moreover, classification of racial and ethnic groups varies considerably in research and practice.

We emphasize at the outset that the problems we consider in this paper are distinct from familiar problems of covariate measurement error (e.g. Dong and Millimet, 2025) and covariate misclassification (e.g. Bollinger 1996, and Molinari 2008). A key distinction is that we do not assume that there are true measures of the covariates of interest. A possible source of semantic confusion is that some research on covariate measurement error uses the word "differential" in a different way than we do. For example, Imai and Yamamoto (2010) write (p. 545): "measurement error in the treatment variable is differential if it is not conditionally independent of outcome given observed covariates. See also Chalak and Kim (2019).

Instead, we allow for the classification of covariates to differ across circumstances with no need for recourse to a "true" measure. We also emphasize here that the paper is concerned primarily with how to use covariate information in predictions, less so with the nature of covariate classification *per se*, although these are intertwined to some degree.

We wrote in the concluding paragraph of Manski et al. (2023):

> "The question is whether the race category that serves as the basis of the clinician's treatment decision…is the same race category that would have been coded for this patient had they been a participant in the clinical trial. If so, the preceding analysis

---

[4] The OMB directives have vast influence on the manner in which race and ethnicity data are collected in federal and other data systems. Changes to Census and other survey questions are not made lightly but are instead based on in-depth empirical study. Interested readers might consult the U.S. Census Bureau's report on the 2015 National Content Test (U.S. Census Bureau, 2017) for further details.

goes through without modification. If not, a more complex analysis must be pursued

that is beyond this paper's scope."

Among other objectives, this paper pursues that more complex analysis.

Section 2 develops our terminology and documents the variation in racial/ethnic classification. Section 3 studies the identification problem, showing how the identifiability of P(y|x) depends on what is known about the relationship between y, w, and x. The present problem of partial identification is broadly similar to, but differs in some details from, those encountered that arise in the closely related contaminated sampling problem (Horowitz and Manski, 1995), mixing problem (Manski, 1997), ecological inference problem (Manski, 2018), and inference on long from short regressions (Cross and Manski, 2002; Li, Litvin, and Manski, 2023). Our primary contribution here is not to perform new identification analysis, but rather to elucidate how the DCC problem -- which has not previously been studied from a partial identification perspective -- creates serious identification problems.

We demonstrate that, in the absence of suitable knowledge restricting the joint distribution of (y, w, x), DCC may result in wide bounds on P(y|x). These wide bounds underscore the significance of the DCC identification problem and the difficulty in solving it: narrowing these bounds requires information that currently may only be available in rare settings but can be collected prospectively alongside classification changes. For example, were OMB to again revise its race/ethnicity classification, it could implement both the old and new scheme in a bridging period, so that the joint distribution of (w, x, y) is known at least during this period. To cope with the identification problem in the absence of bridging data, we suggest application of bounded-variation assumptions, which have previously been used to inform related identification problems in Manski and Pepper (2018) and Li, Litvin, and Manski (2023). Such assumptions narrow bounds on mean regressions in a flexible manner and the degree of narrowing depending on the strength of the assumption imposed.

Section 4 illustrates the breadth of applicability of our main findings. This section includes an empirical example illustrating the wide bounds obtained in two straightforward and policy-relevant settings:

estimating P(y|x) when there is a change to race and ethnicity classification similar to what OMB has proposed and estimating P(y|x) when x represents specific subgroups of a larger group w. We collect new nationally-representative survey data in which respondents provide answers to distinct race and ethnicity classification schemes to inform the first exercise and we use existing survey data for the second exercise.

Our analysis of partial identification of probabilistic predictions conditional on covariates is motivated by the use of such predictions in decision making. However, the problem of decision making under ambiguity generated by the present identification problem is too challenging for adequate analysis in this paper. We leave this for future research, which we expect should yield to algebraic and computational analysis of the type performed recently in Manski (2025). We also leave to future research study of decision making, estimation, and inference with finite-sample data.

## 2. Variation in Racial/Ethnic Classification

Here, we describe how classification of one set of attributes – race and ethnicity – varies across research studies, official statistics, clinical settings, and over time. In doing so, we highlight the importance of DCC in common, policy-relevant settings.

We first outline a general framework to contextualize such variation as a prelude to our analysis of the identification problem occurring when one seeks to learn health outcome probabilities conditional on covariates x, but research evidence only provides estimates of outcome probabilities conditional on w – the DCC problem. We then provide examples of differential classification of race/ethnicity in research studies, official statistics, and in clinician practice environments.

### 2.1. Attributes, Circumstances, and Labels

### 2.1.1. Circumstances

Consider person n in population N. Let C denote a set of circumstances that n could potentially experience. A circumstance c could be a time period, a geographic location, an institutional setting, or a cultural norm. This paper considers circumstances designating patient participation as a subject in a research study or presentation for diagnosis and treatment in a clinic.

### 2.1.2. Attributes

We suppose that person n can be described by a set of potential attribute vectors. $A_n(c)$, $c \in C$, is the attribute vector that would be realized for person n in circumstance c. The components of $A_n(c)$ may or may not be observable and they may or may not vary over c. For example, person n's genome or sex assigned at birth are invariant over c while their age and BMI are not.

A potential attribute vector is akin to a potential outcome vector in familiar analysis of treatment response. In the present context, $A_n(c)$ is realized when person n experiences circumstance c. In principle, person n could be observed in multiple circumstances (e.g., in different clinics, administrative settings, or over time), so that multiple attribute vectors could be realized for this person. This is negligibly likely in the setting we study, so the potential outcome analogy is appropriate. In our setting, the relevant circumstances are denoted c = 1 for participation in a research study and c = 2 for presentation at a clinic for treatment. As such we assume that if circumstance c is realized, then potential attribute vectors $A_n(c')$, $c' \neq c$, are not realized.

### 2.1.3. Labels

Labels are fundamental to our analysis. Labels are the numerical or nominal values that are attached to attributes and that define the covariate values used in empirical studies. How labels are attached to attributes will vary from one circumstance to another, depending in part on the prevailing incentives and constraints that influence labelers at each $c \in C$. However accomplished, attaching labels to attributes is what we call classification.

Let $L_m(c)$ be the set of labels that could be placed on attribute m in circumstance c for each n. The components of label sets may be numerical or nominal. For our setting, the attribute m of interest is race/ethnicity and the relevant label sets are nominal. For example, it may be that $L_m(1) = \{$White, Nonwhite, missing data$\}$ and $L_m(2) = \{$White, Black, Asian, Other, missing data$\}$, in which case the clinician observes a more refined classification than was used in the research study. In terms of the (w, x) notation introduced earlier, $w_n = l_{m,n}(1)$ and $x_n = l_{m,n}(2)$. where $l_{m,n}(c) \in L_m(c)$ is the attribute-m label observed for person n in circumstance c.

Since our focus is mainly on a single attribute and a representative individual, we henceforth drop the m and n subscripts. We assume that the label sets L(c) do not vary with n; i.e. the L(c) are defined by some standards or protocols. Thus, for each n the assigned labels $l(c)$ must be elements of L(c). Central in what follows is the joint distribution across c of labels, $P(l(1), \ldots, l(c^*))$, $l(c) \in L(c)$, where c* is the number of circumstances of interest (c* = 2 for our purposes). Note that commonality of labels across circumstances does not imply commonality of attributes. For example, we may have L(1) = L(2) = \{White, Nonwhite, missing data\}, indicating that the clinician has available the same label set as the research study. However, the data collection process may differ in the two circumstances. If n could experience both c = 1 and c = 2, nothing would preclude $l(1)$ = "White" and $l(2)$ = "Nonwhite" or preclude $l(1)$ = "missing" and $l(2)$ = "Nonwhite". Conversely it may hold that L(1) ≠ L(2) but that  $l(1) = l(2)$ = "Asian" if both L(1) and L(2) include "Asian" as a category.

Without knowledge of the relationship of labels across circumstances, observation of person n's label in one circumstance implies nothing about this n's label in others. Even if an attribute is given the same label across circumstances, that label need not have the same meaning. In general, observation of $w = l(1)$ reveals nothing about x even if the L(1) and L(2) label sets are identical. If label sets vary across circumstances, the potential lack of shared meaning is an obvious concern. Perhaps less appreciated is that identical labels observed across circumstance-invariant label sets may also lack shared meaning.

Formally, assumptions with identifying power may take the form of cross-circumstance restrictions on the population distribution P[y, A(c), c ∈ C], where y is an outcome of prediction interest. Many such assumptions may be asserted, and their implications studied. An important class of assumptions that may sometimes be credible in racial/ethnic classification is cross-circumstance aggregation of attributes. One might assume that each label in L(1) has the same meaning as the union of labels in a non-singleton proper subset of L(2). If so, then label set L(1) would be an informative aggregation or coarsening of L(2). Equivalently, L(2) would be a disaggregation or refinement of L(1).

In the example where L(1) = {White, Nonwhite, missing data} and L(2) = {White, Black, Asian, Other, missing data}, L(1) would be an informative aggregation of L(2) if "White" is assumed to have the same meaning in both circumstances, "Nonwhite" in c = 1 is assumed to have the same meaning as "Black or Asian or Other" in c = 2, and missingness occurs in the same cases in both circumstances. Such assumptions about labels having the same meaning across circumstances imply a type of nominal cross-circumstance aggregation that is sometimes assumed or implied in empirical work. Yet there may be no credible cognitive, behavioral, or empirical foundation for such aggregation, so empirical work based on this sort of superficial aggregation deserves scrutiny. Moreover, while disaggregation (refinement) of categories may in principle offer opportunities to improve identification, such data may not be available in practice. As a practical matter, analysts will often be constrained by the decisions made by each circumstance's category definer.[5]

## 2.2. Racial/Ethnic Classifications in Official Statistics

Having introduced basic ideas abstractly, we now discuss racial/ethnic classification schemes that have been used in practice. We consider three important settings: government statistics, medical research studies, and clinical practice contexts.

The U.S. Office of Management and Budget (OMB) recently revised its longstanding Statistical Policy Directive 15 (SPD-15) that specifies standards for the collection of race and ethnicity information in federal data systems, including the decennial Census and the annual American Community Survey (U.S. OMB, 2024). The new standards eliminate the previous two-question structure in which race and Hispanic ethnicity questions were administered separately. Instead, they add "Hispanic or Latino" and "Middle Eastern or North African" response options to the new single question "What is your race and/or ethnicity?" It is reasonable to anticipate that non-federal data systems (e.g. state administrative data systems; health systems' electronic medical records; etc.) will begin to adopt the new category nomenclature. We discuss some implications of these proposed changes in section 4.

---

[5] It may sometimes be credible to make assumptions like P($l(2)$ = "White" | $l(1)$ = "White") ≥ P($l(2)$ = "White"), i.e. various forms of positive dependence (Lehmann, 1966). While dependence assumptions like this may have identifying power in some instances, an examination of such possibilities is beyond the scope of this paper.

## 2.3. Racial/Ethnic Classifications in Research Studies

In January 2024 the U.S. Food and Drug Administration (FDA) published a Draft Guidance recommending that specific race and ethnicity classifications be used in clinical trials and other clinical studies (U.S. FDA, 2024). Pages 5-6 of the Draft Guidance state:

"For ethnicity, we recommend the following minimum choices be offered: Hispanic or Latino, Not Hispanic or Latino.

"For race, we recommend the following minimum choices be offered: American Indian or Alaska Native, Asian, Black or African American, Native Hawaiian or Other Pacific Islander, White."

The guidance goes on to say that in other situations greater granularity may be preferred and FDA recommends doing so. Anticipating revisions to the U.S. OMB race and ethnicity categories discussed above, the FDA guidance notes plans to revise its recommendations accordingly.

Reflecting changing guidance, classification of patient race and ethnicity in research studies has varied over time and across contemporaneous studies. A review by Turner and colleagues (2022) examines over 20,000 U.S. clinical trials published between 2000 and 2020. Overall, less than a quarter of trials over this period reported all five groups recommended in the FDA draft guidance. However, the proportion doing so increased over time, from 20% prior to 2007 (when the trials database clinicaltrials.gov was launched) to over 40% by 2020. Among the studies reporting any race/ethnicity information, the proportion of trials reporting all five groups was 46%.

The difference in race/ethnicity categories across research studies is also manifest in large survey datasets used for observational research. For example, in 2022, the Behavioral Risk Factor Surveillance Survey (BRFSS) allowed participants to self-identify multiple racial categories -- including White, Black, American Indian or Alaska Native, seven different subgroups of Asian, and four different subgroups of Pacific Islander -- and Hispanic ethnicity. In contrast, the National Health and Nutrition Examination Surveys (NHANES) included only five mutually exclusive options, combining race and ethnicity (Mexican American, Other Hispanic, Non-Hispanic White, Non-Hispanic Black, Non-Hispanic Asian, and other).

## 2.4. Racial/Ethnic Classifications in Clinical Practice

Most clinical documentation in the U.S. occurs within electronic medical record (EMR) systems. Race and ethnicity categories used for U.S. electronic medical records are based on those specified by a standards-setting body called HL7. Health system customers of major EMR vendors (e.g. Epic Systems) can use the HL7 categories in ways that suit their purposes, which may vary across health systems.[6]

Considering how race and ethnicity are recorded within EMR systems reveals heterogeneity across space and over time. In a study examining EMR data from 56 health care institutions, attempts to harmonize data across the ten different reporting schema used across institutions revealed many non-conforming categories (Cook et al, 2022). Missingness of race/ethnicity information was high, particularly for Hispanic patients. High rates of missingness and time-varying classification, as inferred from changes in recorded classification within a single patient, have also been demonstrated in other studies (Agawu et al, 2023).

---

[6] The current HL7 categories for race and ethnicity are found, respectively, at:
https://terminology.hl7.org/2.0.0/CodeSystem-v3-Race.html
and https://build.fhir.org/ig/HL7/UTG/ValueSet-v3-Ethnicity.html.

## 3. Identification Analysis

We now formally investigate the DCC identification problem: drawing conclusions about $P(y|x)$ by combining knowledge of $P(y|w)$ and $P(x)$ with other information that may be available, through some combination of data collection and credible assumptions. Knowing $P(y|w)$ and $P(x)$ allows one to bound $P(y, w, x)$ or $P(y|w, x)$ which then yield bounds on $P(y|x)$. In what follows, we suppose that w and x have finite domains W and X, with $P(w = \omega) > 0$ and $P(x = \xi) > 0$ for all $\omega \in W$ and $\xi \in X$. The outcome y is real-valued. In the notation of section 2.1, W and X correspond to the label sets $L(1)$ and $L(2)$.

The analysis in this section is initially abstract. We also discuss the tractability of computation of derived bounds on the mean $E(y|x)$ and other features of $P(y|x)$. To add concreteness, we sometimes contextualize our analysis in the problem of a clinician seeking to predict (and treat) disease y given observed covariates x, but where only evidence on $P(y|w)$ is available in clinical studies. In Section 4, we show how the DCC problem applies to a range of other settings as well.

An obvious extension is to settings where predictions of y from multiple research studies are available, each study conditioning on different covariates or conditioning on covariates that are nominally identical but that were obtained under different circumstances. Another possibility are research studies that provide information on $P(y_1|x)$, where $y_1$ is an outcome that is related to the outcome of interest $y_0$, either by sharing a common set of underlying drivers (e.g., high-blood pressure or diabetes) or precedes it through some causal mechanism (e.g., diabetes and chronic kidney disease). Then a clinician or other decision maker may wish to learn $P(y|x)$ by combining knowledge obtained from these studies. We do not study such scenarios here, but we believe it important to do so as extensions of our analysis.

Observation of $P(y|w)$ alone reveals nothing about $P(y|x = \xi)$ for a specified $\xi \in X$. Among the types of further information that may be available for identification, we consider restrictions on the distribution $P(w, x)$ in Section 3.1. We consider restrictions that directly relate $P(y|w)$ and $P(y|x)$ in Section 3.2.

### 3.1. Identification with Restrictions on P(w, x)

The central tool when studying identification with restrictions on P(w, x) is to combine two forms of the Law of Total Probability. These are:

(1)  $P(y|w = \omega) \; = \; \sum_{\xi} P(y|w = \omega, x = \xi)P(x = \xi|w = \omega), \; \omega \in W,$

(2)  $P(y|x = \xi) \; = \; \sum_{\omega} P(y|w = \omega, x = \xi)P(w = \omega|x = \xi), \; \xi \in X$

Our core analysis considers the most informative setting, in which P(w, x) is completely known. Following this, we discuss settings in which P(w, x) is partially known.

With knowledge of P(w, x), equation (1) poses a mixture decomposition problem that has been studied in multiple literatures, including those on ecological inference (Cross and Manski, 2002; Manski, 2018) and on contaminated sampling (Horowitz and Manski, 1995). We paraphrase here. We also show how the second form of the Law of Total Probability in equation (2) informs analysis of the DCC problem.

### 3.1.1. Identification in Abstraction

Given knowledge of P(y|w) and P(w, x), the problem in equation (1) is to learn about P(y|w, x). For each $\omega \in W$, this equation restricts [P(y|w = $\omega$, x = $\xi$), $\xi \in X$] to vectors of distributions that solve the equation. There is no information that relates these vectors of distributions across values of w. Hence, the joint identification region for P(y|w, x) is:

(3)  $H[P(y|w, x)] \; = \; [\gamma_{\omega\xi} \in \Gamma_Y, \, \omega \in W, \, \xi \in X: P(y|w = \omega) \; = \; \sum_{\xi \, \in \, X} \gamma_{\omega\xi} \, P(x = \xi|w = \omega)],$

where $\Gamma_Y$ is the space of all distributions on Y. This identification region is always non-empty. In particular, letting P(y|w = $\omega$, x = $\xi$) = P(y|w = $\omega$) for all $\xi \in X$ solves (1).

Given the identification region for P(y|w, x) and the assumed knowledge of P(w|x), equation (2) yields the identification region for P(y|x). The feasible values for these conditional distributions are obtained by inserting all feasible collections of distributions [P(y|w = ω, x = ξ), ω ∈ W, ξ ∈ X] into (2).

In general, P(y|x) is partially identified. An important exception with point identification occurs when x aggregates w (see section 2.1.3). Then there exists a many-to-one function x(·): W → X mapping w into a single value of x. When x aggregates w, P[x = x(ω)|w = ω] = 1, ω ∈ W. Then equations (1) and (2) reduce to:

(1')  P(y|w = ω)  =  P(y|w = ω), ω ∈ W,

(2')  P(y|x = ξ)  =  $\sum\limits_{\omega}$ P(y|w = ω)P(w = ω|x = ξ),   ξ ∈ X.

Thus, (1') is an identity and (2') point-identifies P(y|x).

An important case of partial identification of P(y|x) occurs when w aggregates x. Then there exists a many-to-one function w(·): X → W mapping x into a single value of w. When w aggregates x, P[w = w(ξ)|x = ξ] = 1, ξ ∈ X. Then equations (1) and (2) reduce to:

(1")  P(y|w = ω)  =  $\sum\limits_{\xi}$ P(y|x = ξ)P(x = ξ|w = ω),   ω ∈ W,

(2")  P(y|x = ξ)  =  P(y|x = ξ),   ξ ∈ X.

Now (2") is an identity and (1") partially identifies P(y|x).

When the outcome y is other than binary, the structure of the identification region (3) is generally complex when the dimension of x is greater than two. Cross and Manski (2002) characterized the region for the collection of conditional means E(y|w, x). It is shown that the identification region for [E(y|w = ω,

x = ξ), ξ ∈ X] is a bounded convex set whose extreme points are the expectations of certain |X|-tuples of what they call *stacked distributions*.

The structure of the region simplifies considerably if y is binary, say taking the value 0 or 1. Then the content of (1) and (2) is contained in the linear equations:

(1''')     $P(y = 1 | w = \omega) = \sum_\xi P(y = 1 | w = \omega, x = \xi) P(x = \xi | w = \omega), \omega \in W,$

(2''')     $P(y = 1 | x = \xi) = \sum_\omega P(y = 1 | w = \omega, x = \xi) P(w = \omega | x = \xi), \xi \in X,$

and the linear inequalities $0 \le P(y = 1 | w = \omega, x = \xi) \le 1, \ \omega \in W, \xi \in X$. The sharp bound on $P(y = 1 | x = \xi)$ may be obtained by solving a linear programming problem to minimize and maximize $P(y = 1 | x = \xi)$ subject to the set of linear equations and inequalities.

### 3.1.2. Identification with Binary x

The general characterization of identification simplifies considerably when the observed covariate x is binary, say taking the values 0 and 1. We do not need to restrict the domain of w. This case covers settings with a rudimentary clinical racial/ethnic classification of patients into two categories (e.g., White and non-White).

To simplify the notation, we let $p_\omega \equiv P(x = 0 | w = \omega)$. Then the Law of Total Probability (1) becomes:

(4)     $P(y | w = \omega) = p_\omega P(y | w = \omega, x = 0) + (1 - p_\omega) P(y | w = \omega, x = 1).$

For each value of w, the identification region for $[P(y | w = \omega, x = 0), P(y | w = \omega, x = 1)]$ is:

(5)     $H[P(y | w = \omega, x = 0), P(y | w = \omega, x = 1)] = \{(\gamma_{\omega 0}, \gamma_{\omega 1}) \in \Gamma_Y \times \Gamma_Y : P(y | w = \omega) = p_\omega \gamma_{\omega 0} + (1 - p_\omega) \gamma_{\omega 1}\}.$

Result (5) is simple, but still abstract. The first step in developing its practical implications is to observe that it is enough to study identification of one of the two distributions $P(y|w = \omega, x = 0)$ or $P(y|w = \omega, x = 1)$. This is so because, by (4), specification of one of these distributions implies a unique value for the other. Hence, determination of either $H[P(y|w = \omega, x = 0)]$ or $H[P(y|w = \omega, x = 1)]$ suffices to determine the joint identification region $H[P(y|w = \omega, x = 0), P(y|w = \omega, x = 1)]$.

For specificity, consider $P(y|w = \omega, x = 1)$. Manipulation of (4) yields:

(6)    $P(y|w = \omega, x = 1) = [P(y|w = \omega) - p_\omega P(y|w = \omega, x = 0)]/(1 - p_\omega)$.

Letting $P(y|w = \omega, x = 0)$ range over all elements of $\Gamma_Y$ yields the identification region:

(7)    $H[P(y|w = \omega, x = 1)] = \Gamma_Y \cap \{[P(y|w = \omega) - p_\omega \gamma_{\omega 0}]/(1 - p_\omega), \ \gamma_{\omega 0} \in \Gamma_Y\}$.

*Identification of Event Probabilities*

For further simplification, let B be any subset of Y and consider the event probability $P(y \in B|w = \omega, x = 1)$. The identification region for $P(y \in B|w = \omega, x = 1)$ is the interval

(8)    $H[P(y \in B|w = \omega, x = 1)] = [0, 1] \cap [[P(y \in B|w = \omega) - p_\omega]/(1 - p_\omega), P(y \in B|w = \omega)/(1 - p_\omega)]$.

The lower bound is greater than zero, hence informative, when $p_\omega < P(y \in B|w = \omega)$. The upper bound is less than one, hence informative, when $p_\omega < 1 - P(y \in B|w = \omega)$. When both conditions hold, which requires that $p_\omega$ be less than ½, the interval has width $p_\omega/(1 - p_\omega)$.

The finding in (8) cautions that combination of observation of $P(y|w)$ with knowledge of $P(w, x)$ may not reveal much about $P(y|x)$, indeed sometimes nothing at all. To demonstrate, it suffices to consider the simple case where y, w, and x are all binary, taking the values 0 and 1. We consider identification of $P(y = 1|x = 1)$ for specificity. We continue to use the shorthand $p_\omega \equiv P(x = 0|w = \omega)$.

Knowledge of $P(y|w)$ and $P(w, x)$ is completely uninformative about $P(y = 1|x = 1)$ if $p_\omega \geq \max[P(y = 0|w = \omega), P(y = 1|w = \omega)]$, $\omega \in \{0, 1\}$. Then the joint identification region for $[P(y = 1|w = 0, x = 1), P(y = 1|w = 1, x = 1)]$ is the unit square $[0, 1]^2$. It follows from (2) that the identification region for $P(y = 1|x = 1)$ is $[0, 1]$.

Knowledge of $P(y|w)$ and $P(w, x)$ is informative in other cases, but the implied bound on $P(y = 1|x = 1)$ may be wide. Suppose that $p_\omega < \min[P(y = 0|w = \omega), P(y = 1|w = \omega)]$, $\omega \in \{0, 1\}$. Then the joint identification region for $[P(y = 1|w = 0, x = 1), P(y = 1|w = 1, x = 1)]$ is:

$$[P(y = 1|w = 0) - p_0]/(1 - p_0), P(y = 1|w = 0)/(1 - p_0)]$$
$$\times \; [P(y = 1|w = 1) - p_1]/(1 - p_1), P(y = 1|w = 1)/(1 - p_1)].$$

It now follows from (2) that the identification region for $P(y = 1|x = 1)$ is

$$[P(w = 0|x = 1)[P(y = 1|w = 0) - p_0]/(1 - p_0) + P(w = 1|x = 1)[P(y = 1|w = 1) - p_1]/(1 - p_1)],$$
$$P(w = 0|x = 1)[P(y = 1|w = 0)/(1 - p_0)] + P(w = 1|x = 1)[P(y = 1|w = 1)/(1 - p_1)].$$

This interval has width $P(w = 0|x = 1)p_0/(1 - p_0) + P(w = 1|x = 1)p_1/(1 - p_1)$.

To obtain further insight into the identification problem, suppose that w and x are alternative ways to classify persons into non-White (0) and White (1) categories. Suppose that the definition of "White" is broader in the research study than in the clinical context. Thus, every patient classified as White in the clinic would be classified as White in the research study. Then $P(w = 1|x = 1) = 1$, $P(w = 0|x = 1) = 0$, and the width of the bound on $P(y = 1|x = 1)$ is $p_1/(1 - p_1)$. The quantity $p_1 \equiv P(x = 0|w = 1)$ is the fraction of persons classified non-white in the clinic, among those classified white in the research study.

*Identification of Means and Quantiles*

Considering general outcomes y and features of P(y|x) beyond event probabilities, Horowitz and Manski (1995) derived sharp bounds on the conditional mean E(y|w = ω, x = 1) and α-quantiles $Q_α$(y|w = ω, x = 1), where α ∈ (0, 1). Identification of these parameters can be addressed directly, but it is easier to prove a general result for the class of parameters that respect stochastic dominance and then apply this result to the mean and quantiles. Appendix A of this paper summarizes the analysis.

### 3.1.3. Identification with Partial Knowledge of P(w, x)

Complete knowledge of P(w, x) may sometimes be available through auxiliary data collection that measures patient attributes in multiple ways, or such knowledge may be asserted by posing assumptions. However, it may be that P(w, x) is only partially known. If so, the identification region for P(y|x) is necessarily a superset of the one derived above.

Abstractly, one may know that P(w, x) lies in some set H[P(w, x)] of bivariate distributions with domain W × X. A salient case is one in which a research study reports P(w) and clinical practice reveals P(x). Then data collection only yields knowledge of the marginals of P(w, x). Research on Boole-Frechét-Hoeffding bounds has characterized the resulting identification region for P(w, x) in the absence of assumptions restricting this distribution.

Whatever form the set H[P(w, x)] takes, the identification region for P(y|x) can be derived abstractly by extension of the analysis in Section 3.1.1. We explain further in Appendix B. We should caution that computation of sharp bounds may be complex even when the outcome y is binary. If the conditional probabilities P(w|x) and P(x|w) are only partially known, equations (1''') and (2''') are bilinear rather that linear in the unknown quantities. Solution of bilinear programming problems is often challenging. See the discussion of computation in Li, Litvin, and Manski (2023), whose identification problem had a bilinear structure.

## 3.2. Identification with Assumptions Relating P(y|w) and P(y|x)

The main lesson of the analysis in Section 3.1 is that knowledge of P(y|w) and P(w, x) often does not suffice to reveal much about P(y|x). Narrower bounds may be obtained by adding assumptions that directly relate P(y|w) and P(y|x).

Particularly simple are *bounded-variation* assumptions, which constrain the distance between the two distributions. In settings that differ substantively from DCC but are mathematically similar, such assumptions have been applied by Manski (2018), Manski and Pepper (2018), Li, Litvin, and Manski (2023), and elsewhere. When y, w, and x are all binary, a leading example of a bounded-variation assumption is to assert that $|P(y = 1|w = 0) - P(y = 1|x = 0)| < \delta_0$ and $|P(y = 1|w = 1) - P(y = 1|x = 1)| < \delta_1$, where $\delta_0$ and $\delta_1$ are specified positive constants.

Abstract identification analysis with bounded-variation assumptions is generally straightforward. The tighter the constraint on the distance between P(y|w) and P(y|x), the greater is the identifying power. However, assessment of the credibility of such an assumption may not be straightforward. Credibility necessarily is context-specific.

When outcome y is binary and P(w, x) is completely known, computation of the sharp bound on P(y = 1|x) is tractable. Bounded variation assumptions impose linear inequalities on the unknown conditional probabilities. Hence, sharp bounds are obtained by combining them with (1''') and (2''') and solving linear programming problems.

## 4. Applications

While we examined a clinical decision problem to fix ideas around identification challenges given DCC in section 3, our analysis generalizes to many different applications. We discuss and empirically illustrate a policy relevant application in Section 4.1: harmonization and bridging across different classifications of race and ethnicity. Section 4.2 provides an empirical illustration of partial identification of race/ethnicity specific disease risks for clinically relevant subcategories in situations where P(y|w) is

only available for the larger category comprising these subgroups.

4.1. Harmonization and Bridging

The recent update of OMB's SPD-15 (U.S. OMB, 2024) specifies new race and ethnicity data collection standards to be used across federal census and survey systems. Most prominent in the revision is the replacement of the earlier two-question race (five categories) and ethnicity (two categories) structure with a single-question, seven-category structure whose simplest form uses: American Indian or Alaska Native; Asian; Black or African American; Hispanic or Latino; Middle Eastern or North African; Native Hawaiian or Pacific Islander; and White (allowing for multiple-category responses).

One concern is how to reconcile or harmonize statistical findings that were based on the pre-revision categories with those now to be obtained with the new categories. OMB refers to such harmonization as "bridging" (U.S. OMB, n.d.). In this paper's terminology, the pre- and post-revision periods can be viewed as two distinct circumstances in which covariates w and x are available, respectively. Having learned P(y|w) over a time series of pre-revision data, how can these probabilities be compared with post-revision findings when only y and x are available?

We illustrate the partial identification challenges introduced by changes in race and ethnicity classification schema. Specifically, we evaluate a shift from the two-question standard to the newly proposed single question standard described above. To do so, we conducted a survey in October 2025, introducing the latter question to 955 respondents ages 18-64 who are members of the YouGov nationally representative longitudinal panel. These panelists are asked to provide answers to the two-question standard annually. To reduce the potential of anchoring effects, we fielded our survey to respondents who had last answered questions on race and ethnicity at least 3 months prior to our survey. We also asked participants to report whether they had ever been diagnosed with diabetes or hypertension.

The distribution of responses to questions using the current and proposed OMB standards are presented in Table 1. We aggregate responses to the recently proposed OMB standard to be comparable to the existing standard. The table reveals that 12% of respondents provide distinct answers to the two classification schemes. In Appendix C, we bound the prevalence of diabetes among individuals reporting Hispanic ethnicity under the new OMB scheme, $P(y = 1, x = 1)$, when only $P(w = 1, x = 1)$ and $P(y = 1, w = 1)$ are known. The bound is [0.09, 0.39], whereas the actual value for $P(y = 1, x = 1)$ based on the survey data is 0.136. To narrow the bound would require that we maintain assumptions relating $P(y|w)$ and $P(y|x)$.

U.S. OMB (n.d.) describes various bridging strategies that federal agencies might use. These strategies are designed to provide researchers knowledge of $P(w = 1, x = 1)$. For example, they propose crosswalks to map race/ethnicity across categories, many of which involve weights ("bridging factors") to achieve 1:1 categorizations. However, as we have seen in our empirical example, even full knowledge of $P(w = 1, x = 1)$ does not address the partial-identification nature of the fundamental problem these agencies confront. This problem remains to learn when $P(y|x)$, when historical data only reveal $P(y|w)$. More generally, as we noted in section 2.1.3, superficial aggregation does not generally resolve identification in bridging exercises. Millimet (2024) reports that bridging when categories do not superficially aggregate is no less problematic, inducing "measurement error that is necessarily nonclassical and can be consequential."

## 4.2. Diabetes Risk Within Asian Subgroups

An ongoing controversy in clinical medicine is the lack of research data on disease risks and treatment outcomes among subgroups of individuals categorized as Asian. This group comprises of a range of individuals who originate or descend from a range of countries, e.g., China, India, Japan, the Philippines, and may thus be expected to have differing disease risks. Early evidence of heterogeneity across these groups (e.g., Cheng et al, 2019; Lui et al, 2024) has prompted calls for updated race/ethnicity classification that disaggregates the Asian group into these relevant subgroups (Yi, 2024). Neither the standard nor the proposed OMB classification scheme addresses the need to disaggregate the Asian group.

In the absence of widespread data collection of this nature, one may apply the methods developed in Section 3.1.2. To do so, the researcher or clinician needs information on $P(y|w)$, where $w$ represents the aggregated Asian category, and the probability that a patient classified as Asian belongs to the subgroup of interest. We conduct this exercise using data from the 1999-2018 National Health Interview Surveys (NHIS), which includes information for 25,500 adult Asian respondents classified into four mutually exclusive subgroups: Chinese, Filipino, Asian Indian, and Other Asian.[7] To connect with the clinical literature on this topic (Chang et al 2019), we focus on the outcome of whether the respondent has diabetes (y). In our sample, P(y = 1|w) = 0.0898. Among the Asian respondents, the proportion Chinese is 0.2284; the proportion Filipino 0.1975; the proportion Asian Indian 0.2208; the proportion reporting another subgroup 0.3603.

Using these quantities and applying Equation (8), we can obtain the following bounds for reporting a diabetes diagnosis conditional each Asian subgroup:

P(y = 1|x = Chinese) ∈ [0, 1] ∩ [(0.0898 - 0.7786)/0.2284, 0.0898/0.2284] = [0, 0.3932]

P(y = 1|x = Filipino) ∈ [0, 1] ∩ [(0.0898 - 0.8025)/0.1975, 0.0898/0.1975] = [0, 0.4547]

P(y = 1|x = Asian Indian) ∈ [0, 1] ∩ [(0.0898 - 0.7792)/0.2208, 0.0898/0.2208] = [0, 0.4067]

P(y = 1|x = other Asian) ∈ [0, 1] ∩ [(0.0898 - 0.6397)/0.3603, 0.0898/0.3603] = [0, 0.2492]

Since the Asian subgroups are known in our sample, we can compare these bounds to the point estimates:

P(y = 1|x = Chinese) = 0.0638

P(y = 1|x = Filipino) = 0.1003

P(y = 1|x = Asian Indian) = 0.1118

P(y = 1|x = other Asian) = 0.0842

The wide bounds, derived making no assumptions about the distribution of diabetes across Asian subgroups, underscore the potential gravity of the differential classification problem. The upper bound for each subgroup holds if all diabetes is concentrated in this subgroup. Accordingly, the diabetes rate in the other subgroups are all zero, their lower bounds.

---

[7] Specifically, we regress a binary indicator denoting whether the respondent had ever been diagnosed with diabetes on a vector of fixed effects for racial group and age-gender fixed effects. We exclude individuals denoting Hispanic ethnicity from the sample for simplicity, though the results are similar when we include this group. We account for complex survey design by weighting using the NHIS sampling weights. Our regression allows us to recover age and gender-adjusted prevalences of diabetes for each racial group.

One may narrow the bounds by making bounded variation assumptions that restrict the risks of diabetes. For example, one might assume that the risk in each Asian subgroup must lie between the lowest and highest rates of diabetes estimated in epidemiologic studies of other racial and ethnic subgroups. Even with this assumption, one may not be able to do much to discern the magnitude of differential risks across groups. For example, one may not be able to discern whether Asian Indians have a two-fold higher prevalence of diagnosed diabetes than Chinese or half the Chinese prevalence. Clinicians would like to have refined knowledge of diabetes risks to accurately guide diagnostic testing and treatment.

## 5. Discussion

Data collection involves making choices on what to measure and how to measure it. Changes in how particular attributes are classified have long come with changes in social norms, demographic patterns, and policy goals. While the reasons behind these shifts may be laudable, differential covariate classification introduces a difficult identification problem for researchers examining differences in outcomes across groups (over time or space) and decision-makers seeking to intervene or allocate resources based on risk of some outcome, which varies by demographic characteristics, using existing evidence-based research findings. This identification problem has not received much attention to date. However, it is consequential and relevant. Our analysis reveals that bounds on the object of interest, $P(y|x)$, given knowledge of $P(y|w)$ and $P(w, x)$, may be quite wide. This is in a simple setting where challenges to external validity are assumed away. Narrowing these bounds requires information that is not typically available or potentially untenable assumptions.

One way to address this challenge is to prospectively collect data on $(y, x, w)$ using both classification schemes $w$ and $x$ for a period of time, before switching to a sole use of a single one of these schemes. Doing so would enable estimation of the joint distribution, P(y, x, w). This would solve the identification problem, subject to the assumption that the joint distribution does not change over time. Our questioning of members of the YouGov panel illustrates this type of data collection.

Another way is to identify other related outcomes for which P(y|w) and P(x|w) are both known and use that knowledge to narrow bounds. The returns to all these strategies will need to be evaluated against costs on a case-by-case basis. If evidence-based decision-making is to be conducted rigorously and taken seriously (see Baicker and Chandra, 2017) such evaluation is essential. Our analysis in this paper provides a framework for such discussions.

References

Agawu, A., Chaiyachati, B.H., Radack, J., Duncan, A.F., and A. Ellison. 2023. "Patterns of Change in Race Category in the Electronic Health Record of a Pediatric Population." *JAMA Pediatrics* 177: 536-539.

Antman, F. and B. Duncan. 2015. "Incentives to Identify: Racial Identity in the Age of Affirmative Action." *Review of Economics and Statistics* 97: 710-713.

Baicker K. and A. Chandra. 2017. Evidence-Based Health Policy. *New England Journal of Medicine* 377: 2413-2415.

Baron, E.J., Doyle, J.J., Emanuel, N., Hull, P., and J. Ryan. 2024. "Unwarranted Disparity in High-Stakes Decisions: Race Measurement and Policy Responses." *NBER Working Paper* No. 33104.

Bollinger, C.R. 1996. "Bounding Mean Regressions When a Binary Regressor is Mismeasured." *Journal of Econometrics* 73: 387-399.

Chalak, K. and D. Kim. 2019. "Measurement Error Without the Proxy Exclusion Restriction." *Journal of Business & Economic Statistics* 39: 200–216.

Chang, Y.J., Kanaya, A.M., Araneta, M.R.G., Saydah, S.H., Kahn, H.S., Gregg, E.W., Fujimoto, W.Y., and G. Imperatore. 2019. "Prevalence of Diabetes by Race and Ethnicity in the United States, 2011-2016." *JAMA* 322: 2389-2398.

Charles, K.K. and J. Guryan. 2011. "Studying Discrimination: Fundamental Challenges and Recent Progress." *Annual Review of Economics* 3: 479-511.

Cross, P. and C.F. Manski. 2002. "Regressions, Short and Long." *Econometrica* 70: 357-368.

Cook, L., Espinoza, J., Weiskopf, N.G., Mathews, N., Dorr, D.A., Gonzales, K.L., Wilcox, A., Madlock-Brown, C., for the N3C Consortium. 2022. Issues with Variability in Electronic Health Record Data About Race and Ethnicity: Descriptive Analysis of the National COVID Cohort Collaborative Data Enclave. *JMIR Medical Informatics* 10: e39235.

Davenport, L. 2020. "The Fluidity of Racial Classifications." *Annual Review of Political Science* 23: 221-240.

Degtiar, I. and S. Rose. 2023. "A Review of Generalizability and Transportability." *Annual Review of Statistics and Its Application* 10: 501-24.

Dong, H. and D.L. Millimet. 2025. "Embrace the Noise: It is OK to Ignore Measurement Error in a Covariate, Sometimes." *JRSS-A* (in press).

Duncan, O.D. and B. Davis. 1953. "An Alternative to Ecological Correlation." *American Sociological Review* 18: 665–666.

Finlay, K., Luh, E., and M.G. Mueller-Smith. 2024. "Race and Ethnicity (Mis)measurement in the U.S. Criminal Justice System." *NBER Working Paper* No. 32657.

Horowitz, J. and C.F. Manski. 1995. "Identification and Robustness with Contaminated and Corrupted Data." *Econometrica* 63: 281-302.

Imai, K. and T. Yamamoto. 2010. "Causal Inference with Differential Measurement Error: Nonparametric Identification and Sensitivity Analysis." *American Political Science Review* 54: 543-560.

Lehmann, E.L. 1966. "Some Concepts of Dependence." *Annals of Mathematical Statistics* 37: 1137-1153.

Li, S., V. Litvin, and C.F. Manski. 2023. "Partial Identification of Personalized Treatment Response with Trial-reported Analyses of Binary Subgroups." *Epidemiology* 34: 319-324.

Lui, C.K., Ye, Y., Gee, J., Cook, W.K., Tam, C.C., Sun, S., Miranda, R., Subica, A., and N. Mulia. 2024. "Unmasking Suicidal Ideation for Asian American, Native Hawaiian, and Pacific Islander Youths Via Data Disaggregation." *JAMA Network Open* 7: e2446832.

Manski, C.F. 1997. "The Mixing Problem in Programme Evaluation." *Review of Economic Studies* 64: 537-553.

Manski, C.F. 2013. *Public Policy in an Uncertain World*. Harvard University Press.

Manski, C.F. 2018. "Credible Ecological Inference for Medical Decisions with Personalized Risk Assessment." *Quantitative Economics* 9: 541-569.

Manski, C.F. 2019. *Patient Care under Uncertainty.* Princeton University Press.

Manski, C.F. 2025. "Using Limited Trial Evidence to Choose Treatment Dosage when Efficacy and Toxicity Weakly Increase with Dose." *Epidemiology* 36: 60-65.

Manski C.F., J. Mullahy, and A.S. Venkataramani. 2023. "Using Measures of Race to Make Clinical Predictions: Decision Making, Patient Health, and Fairness." *PNAS* 120(35):e2303370120.

Manski, C. and J. Pepper. 2018. "How Do Right-to-Carry Laws Affect Crime Rates? Coping with Ambiguity Using Bounded-Variation Assumptions," *Review of Economics and Statistics*, 100, 232-244.

Millimet, D. (2024). "(Don't) Walk This Way: The Econometrics of Crosswalks." *IZA Discussion Paper* No. 17154.

Molinari, F. 2008. "Partial Identification of Probability Distributions with Misclassified Data." *Journal of Econometrics* 144: 81-117.

Pan American Health Organization (PAHO). 2021. *Introduction to Semantic Interoperability*. PAHO Document PAHO/EIH/IS/21-023. https://iris.paho.org/handle/10665.2/55417 (retrieved May 30, 2024).

Penner, A.M. and A. Saperstein. 2008. "How Social Status Shapes Race." *PNAS* 105: 19628-19630.

Salhi, R.A. et al. 2024. "Frequency of Discordant Documentation of Patient Race and Ethnicity." *JAMA Network Open* 7: e240549. doi:10.1001/jamanetworkopen.2024.0549.

Saperstein, A. and A.M. Penner. 2012. "Racial Fluidity and Inequality in the United States." *American Journal of Sociology* 118: 676-727.

Turner, B.E., Steinberg, J.R., Weeks, B.T., Rodriguez, F., and M.R. Cullen. 2022. "Race/Ethnicity Reporting and Representation in U.S. Clinical Trials: A Cohort Study." *Lancet Regional Health – Americas* 11: e100252

U.S. Food and Drug Administration. 2024. "Collection of Race and Ethnicity Data in Clinical Trials and Clinical Studies for FDA-Regulated Medical Products." Draft Guidance for Industry. *Federal Register*, January 30, 2024: 5911-5913.

U.S. Office of Management and Budget. 2024. "Revisions to OMB's Statistical Policy Directive No. 15: Standards for Maintaining, Collecting, and Presenting Federal Data on Race and Ethnicity." *Federal Register*, March 29, 2024: 22182-22196.

U.S. Office of Management and Budget (n.d.). *Federal Interagency Technical Working Group on Race and Ethnicity Standards: Annex 6. Bridging Team Methods Report*. https://www2.census.gov/about/ombraceethnicityitwg/annex-6-itwg-bridging-team-methods-report.pdf (retrieved June 3, 2024).

Yi, S. 2024. Data Equity and Multiracial and Multiethnic Communities. *JAMA Network Open* 7: e2446839.

**Table 1**. Distribution of responses to existing and proposed OMB race/ethnicity classification standards

| New OMB | Existing OMB | | | | | | |
|---|---|---|---|---|---|---|---|
| | *NH White* | *NH Black* | *Hispanic* | *Asian* | *Native* | *Mixed/Other* | *Total* |
| *NH White* | 578 | 1 | 2 | 0 | 2 | 15 | 598 |
| *NH Black* | 1 | 94 | 0 | 0 | 0 | 9 | 104 |
| *Hispanic* | 17 | 4 | 110 | 0 | 2 | 21 | 154 |
| *Asian* | 0 | 0 | 0 | 31 | 0 | 17 | 48 |
| *Native* | 6 | 1 | 0 | 0 | 8 | 7 | 22 |
| *Mixed/Other* | 7 | 1 | 0 | 0 | 1 | 20 | 29 |
| *Total* | 609 | 101 | 112 | 31 | 13 | 89 | 955 |

Notes: Data obtained from an original October 2025 survey of 955 respondents ages 18-65 on YouGov panel. Existing OMB refers to the current two question race/ethnicity classification scheme, where respondents are queried about Hispanic ethnicity first and then asked to self-classify in one or more of the following racial categories: American Indian or Alaskan Native, Asian or Pacific Islander, Black, White, Other. New OMB refers to the 2024 proposed standard combining race and ethnicity into a single question, allowing respondents to select one or more of the following: American Indian or Alaska Native, Asian, Black or African American, Hispanic or Latino, Middle Eastern or North African, Native Hawaiian or Pacific Islander, White. NH White and NH Black refer to Non-Hispanic White and Black, respectively. Responses are aggregated to the recently proposed OMB standard to be comparable to the existing standard. The Middle Eastern and North African category is included in "Mixed/Other," as are responses from participants who mark multiple categories.

Horowitz and Manski (1995) showed that the identification region $H[P(y|w = \omega, x = 1)]$ contains a "smallest" member $L_{p\omega}$ that is stochastically dominated by all feasible values of $P(y|w = \omega, x = 1)$ and a "largest" member $U_{p\omega}$ that stochastically dominates all feasible values of $P(y|w = \omega, x = 1)$. These smallest and largest distributions are truncated versions of the distribution $P(y|w = \omega)$: $L_{p\omega}$ right-truncates $P(y|w = \omega)$ at its $(1 - p_\omega)$−quantile and $U_{p\omega}$ left-truncates $P(y|w = \omega)$ at its $p_\omega$−quantile.

With this as background, it follows immediately that if $D(\cdot)$ is a parameter that respects stochastic dominance, the smallest feasible value of $D[P(y|w = \omega, x = 1)]$ is $D(L_{p\omega})$ and the largest feasible value is $D(U_{p\omega})$. Thus, sharp bounds on $E(y|w = \omega, x = 1)$ are the means of $L_{p\omega}$ and $U_{p\omega}$. Sharp bounds on $Q_\alpha(y|w = \omega, x = 1)$ are the $\alpha$−quantiles of $L_{p\omega}$ and $U_{p\omega}$.

The above result determines sharp lower and upper bounds on $D[P(y|w = \omega, x = 1)]$, but it does not assert that the identification region is the entire interval connecting these bounds. The identification region is the entire interval if $D(\cdot)$ is the expectation parameter. However, the interior of the interval may contain non-feasible values if $D(\cdot)$ is a different parameter that respects stochastic dominance. Suppose, for example, that y is discrete and $D(\cdot)$ is a quantile. Quantiles must be elements of the set Y of logically possible values of y. Hence, the identification region for $D[P(y|w = \omega, x = 1)]$ can only include the elements of the interval $[D(L_{p\omega}), D(U_{p\omega})]$ that belong to Y.

Appendix B: Identification with Partial Knowledge of P(w, x)

Let us relabel the identification region for $P(y|w, x)$ given in equation (3) by making explicit its dependence on $P(w, x)$. Thus, let region (3) now be denoted $H_{P(w, x)}[P(y|w, x)]$. Knowing only that $P(w, x) \in H[P(w, x)]$, it follows immediately that

P(y|w, x) $\in$ $\bigcup\limits_{\gamma \,\in\, H[P(w,\, x)]}$ $H_\gamma[P(y|w, x)].$

The right-hand side is the identification region for P(y|w, x) with partial knowledge of P(w, x), specifically with partial knowledge of P(x|w).

Given the above, application of equation (2) across the jointly feasible values of P(y|w, x) and P(w|x) yields the identification region for P(y|x). The simplicity of this abstract statement conceals the possible difficulty of computing the region. The possible difficulty is indicated in the words "jointly feasible values of P(y|w, x) and P(w|x)." A value of P(y|w, x) can be feasible only if it is compatible with some feasible value of P(w|x). The joint identification region for P(y|w, x) and P(w|x) may be a proper subset of the Cartesian Product of their separate identification regions, the specifics depending on the setting. Determination of this joint identification region is necessary to use equation (2) to determine H[P(y|x)].

<u>Appendix C: Bounds on Diabetes Prevalence, P(y = 1|x = 1), When P(w = 1, x = 1) and P(y = 1, w = 1) are known.</u>

Let y, w, and x all be binary, taking the values 0 and 1. Consider identification of P(y = 1|x = 1). In the numerical application, y = 1 means having diabetes. w = 1 means being classified as Hispanic in old OMB and x = 1 means Hispanic in new OMB. w = 0 or x = 0 aggregates non-Hispanic classifications. Recall that

(2)   P(y = 1|x = 1)  =  P(y = 1|w = 0, x = 1)P(w = 0|x = 1) + P(y = 1|w = 1, x = 1)P(w = 1|x = 1).

In the survey data we collected:

P(w = 0|x = 1) = 44/154 = 0.286

P(w = 1|x = 1) = 110/154 = 0.714

P(x = 0|w = 0) = (955 − 112 − 44)/(955 − 112) = 0.948

P(x = 0|w = 1) = 2/112 = 0.018

P(y = 1|w = 1) = 16/112 = 0.143

P(y = 1|x = 1) = 21/154 = 0.136.

P(x = 0|w = 0) > max[P(y = 0|w = 0), P(y = 1|w = 0)]. Hence, the bound on P(y = 1|w = 0, x = 1) is [0, 1].

P(x = 0|w = 1) <  min[P(y = 0|w = 1), P(y = 1|w = 1)]. Hence, the bound on P(y = 1|w = 1, x = 1) is

P(y = 1|w = 1) − P(x = 0|w = 1)]/ P(x = 1|w = 1), P(y = 1|w = 1)/ P(x = 1|w = 1)

        = [(0.143 − 0.018)/0.982, 0.143/0.982]  =  [0.127, 0.146].

Hence, by (2), the bound on P(y = 1|x = 1) is [0 + 0.127·0.714, 1·0.286 +  0.146·0.714] = [0.090, 0.390].

The survey data shows that the actual value of P(y = 1|x = 1) is 0.136.